\documentclass[usenatbib]{mn2e}
\usepackage{epsf}
\usepackage{epsfig}
\usepackage{deluxetable}

\newcommand\fig{{ Fig.}}
\newcommand\figs{{ Figs.}}
\newcommand\tab{{ Table}}
\newcommand\eq{{ Eq.}}
\newcommand\eqs{{ Eqs.}}

\title[Self-Gravitating System Embedded in a Potential Well]{Equilibrium and Dynamical Evolution of Self-Gravitating System Embedded in a Potential Well}

\author[I. Yoon, H. M. Lee, \& J. Hong]{Ilsang Yoon$^{1}$\thanks{e-mail: iyoon@astro.umass.edu},
Hyung Mok Lee$^2$\thanks{e-mail: hmlee@snu.ac.kr}, and
Jongsuk Hong$^2$\thanks{e-maill: chrnodia@astro.snu.ac.kr}\\
$^1$ Department of Astronomy, University of Massachusetts, Amherst, MA 01003, USA\\
$^2$ Astronomy Program, Department of Physics and Astronomy, Seoul National University, 
Seoul 151-747, Korea}

\begin{document}
\maketitle

\begin{abstract}
Isothermal and self-gravitating systems bound by non-conducting and conducting walls 
are known to be unstable if the density contrast between the center and the boundary 
exceeds critical values. We investigate the equilibrium and dynamical evolution of isothermal and self-gravitating system 
embedded in potential well, which can be the situation of many astrophysical objects such as the central parts of the galaxies, or clusters of galaxies with potential dominated by dark matter, but is still limited to 
the case where the potential well is fixed during the evolution.
As the ratio between the depth of surrounding potential well and potential of embedded system becomes large, the potential well becomes effectively the same boundary condition as conducting wall, which 
behaves like a thermal heat bath. 
We also use the direct $N$-body simulation code, NBODY6 to simulate the dynamical evolution of stellar system 
embedded in potential wells and propose the equilibrium models for this system. 
In deep potential well, which is analogous to the heat bath 
with high temperature, the embedded self-gravitating system is dynamically hot, and loosely bound or can be 
unbound since the kinetic energy increases due to the heating by the potential well. 
On the other hand, the system undergoes core collapse by self-gravity when potential well is shallow.
Binary heating can stop the collapse and leads to the expansion, but the evolution is very slow 
because the potential as a heat bath can absorb the energy generated by the binaries. 
The system can be regarded as quasi-static. 
Density and velocity dispersion profiles from the $N$-body simulations 
in the final quasi-equilibrium state are similar to our equilibrium models assumed to be in thermal equilibrium 
with the potential well.
\end{abstract}



\section{Introduction}
Thermodynamics of self-gravitating system is interesting subject but not clearly established yet.
As \citet{padma1990} mentioned, it is probably only fair to say that we do not have a systematic understanding of the 
self-gravitating system at a level similar to the kinetic theory of plasmas.
Although some results from statistical mechanics may not be direclty applied to systems with long-range forces like gravity \citep{bt2008}, self-gravitating systems are correctly described by standard statistical mechanics provided that the thermodynamical limit is correctly defined \citep{padma1990,katz2003, chavanis2006b}.
  
There have been numerous attempts to understand the thermodynamical behavior of self-gravitating
system. Among them, related with stellar dynamics, \citet{antonov1962} studied
the entropy of self-gravitating, isothermal gaseous system surrounded by adiabatic
rigid wall (i.e. non-conducting wall) and found that when the central
concentration exceeds the critical value,
the system can not have a local maximum value of entropy and leads to
runaway instability. 
This is called Antonov's problem suggesting core collapse of stellar system.
\citet{lw1968} extended Antonov's problem to various boundary conditions 
and named the gravothermal catastrophe for this instability, which 
has been confirmed by many analytical and numerical works \citep{hk1978,hs1978,inagaki1980,cohn1980,le1980,jrpz2000}.

The stability of isothermal self-gravitating system was first rigorously investigated by \citet{katz1978}, and later 
reconsidered by \citet{padma1989} using the second variation of entropy. These analyses were 
done in the microcanonical ensemble where the energy of the system is conserved (i.e. non-conducting wall). 
Recently \citet{chavanis2002a} extended the work of Padmanabhan to the canonical ensemble 
where the temperature of the system is fixed (i.e. conducting wall), using the second variation of the free energy.

Since the canonical distribution cannot be derived from the microcanonical distribution in the
presence of long-range interactions \citep{padma1990}, mean field theory has been used to study the thermodynamics of 
self-gravitating systems. In this perspective, self-gravitating isothermal system is stable only if the system is 
in a local maximum of an appropriate thermodynamical potential (i.e. the entropy in the microcanonical ensemble and the 
free energy in the canonical ensemble) as mentioned by \citet{chavanis2002a}.
However \citet{devega2002a,devega2002b} found the `dilute'
thermodynamic limit (particle number $N \rightarrow \infty$ and volume $V \rightarrow \infty$, keeping $N/V^{1/3}$ constant)
where energy, entropy, the free energy are extensive. 
Their works justify the previous analyses and specify the range of validity.

Previous studies of thermodynamical description of self-gravitating system have mostly considered the systems 
enclosed by rigid wall to prevent particle evaporation, which can be justified if a quasi-stationary condition is satisfied 
(i.e. particle evaporation rate is small). \citet{vg2003} replaced this rigid wall by tidal energy prescription 
and investigated Antonov's problem in alternative point of view, which naturally determines the size of the system in addition to 
confirming main features of the isothermal sphere model (i.e. core collapse and negative heat capacity).

Here we propose another natural boundary condition: a potential well which keeps particles 
from evaporating. The completely isolated system is hard to find in astronomy and this type of boundary condition is often
seen in different astronomical scale: for example, cluster galaxies embedded in dark matter potential well
and dense stellar system in galactic nuclei surrounded by much larger bulge.
However the system embedded in potential well and the effect of potential well to the evolution of central 
self-gravitating system have not been studied in thermodynamical point of view.
Therefore in this work, by introducing simple model to describe the 
self-gravitating system in potential well, we attempt to answer the following questions: 
what the role of potential well is, how the embedded system evolves dynamically and what the equilibrium configuration for 
this embedded self-gravitating system can be.
 
This paper is organized as follows.
In Section 2, the previous works on the self-gravitating isothermal sphere surrounded by spherical rigid wall 
are briefly overviewed. We assign this separate section to describe the derivation of some formulae 
and provide interpretations of the previously known results because more detailed description of these previous works is 
necessary to describe our work which replaces the rigid wall by a potential well. 
Then in Section 3, we consider the potential well as a new boundary and study the role of
potential well. In Section 4, we numerically simulate the dynamical evolution of self-gravitating system in potential well. 
The equilibrium models are presented in Section 5. 
In the last section, the results are summarized, and implications and limitations of this study are discussed.

\section{Self-gravitating isothermal sphere enclosed by non-conducting and conducting wall: overview}
Previous analyses on the self-gravitating isothermal sphere surrounded by thermally conducting and non-conducting wall are studied in detail by \citet{padma1989,padma1990}, \citet{katz2003}, 
and \citet{chavanis2002a,chavanis2006b} and
summarized in \citet{bt2008}.
A system composed of $N$ particles can be represented by one particle distribution function $f=f(x,p,t)$ \citep{bt2008}.
With the definition of Boltzmann-Gibbs entropy:
$S \equiv - \int f \ln f d^3 x d^3 p$, the solution with extreme entropy
is well known to be a spherically symmetric isothermal sphere \citep{padma1989,padma1990}.
If we introduce the length, mass and energy scale following \citet{padma1989}:
\begin{eqnarray}
L_0 \equiv (4\pi G \rho_c \beta)^{1/2}, \quad M_0 = 4 \pi \rho_c L_0^3
, \nonumber \\
\phi_0 \equiv \beta^{-1} = \frac{k_B T}{m} = \frac{GM_0}{L_0},
\label{scalevar}
\end{eqnarray} 
and use new dimensionless variables:
\begin{equation}
\xi \equiv \frac{r}{L_0}, \quad n \equiv \frac{\rho(r)}{\rho_c}, \quad m \equiv
\frac{M(r)}{M_0}, \quad \psi \equiv \beta(\phi-\phi_0)
\label{dimlessvar},
\end{equation}
these dimensionless variables satisfy the following relations
\begin{equation}
\psi' = \frac{m}{\xi^2}, \quad m' = n \xi^2, \quad n' = - \frac{mn}{\xi^2}.
\label{dimlessvar2}
\end{equation}
Hereafter, $'$ symbol means the derivative with respect to $\xi$.
Using these relations, we obtain the Lane-Emden equation
\begin{equation}
\frac{1}{\xi^2} \frac{d}{d\xi} \big( \xi^2 \frac{d\psi}{d\xi} \big) = e^{-\psi}
\end{equation}
with the boundary condition $\psi(0)= \psi'(0)=0$. 
Using homology invariants:
\begin{eqnarray}
v &=& \xi \psi' = m/\xi \label{vvar} \\
u &=& \frac{\xi e^{\psi}}{\psi'} = \frac{n\xi^3}{m}, \label{uvar} 
\end{eqnarray}
the equation describing isothermal sphere is transformed to the following coupled differential equations \citep{padma1989}.
\begin{eqnarray}
\frac{1}{u} \frac{du}{d\xi} & = &  \frac{1}{\xi} (3-u-v) \label{uveq1} \\
\frac{1}{v} \frac{dv}{d\xi} & = &  \frac{1}{\xi} (u-1) \label{uveq2} .
\end{eqnarray}
We combine these equations to get
\begin{equation}
\frac{u}{v} \frac{dv}{du}  =  - \frac{u-1}{u+v-3}.
\label{uvplane1}
\end{equation}

\begin{figure}
\center
\epsfig{figure=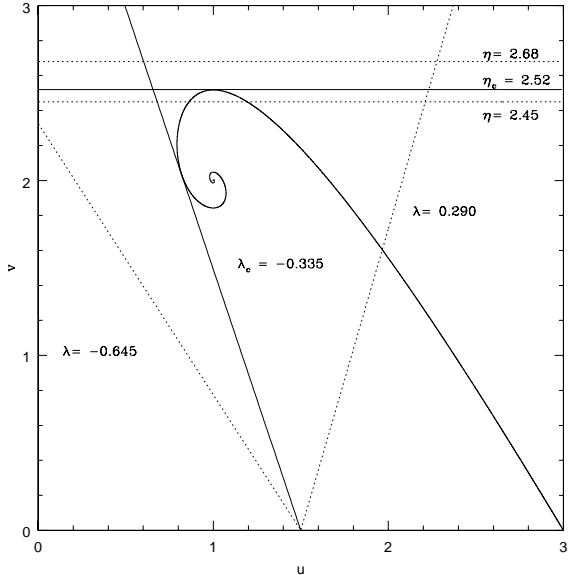, height=0.47\textwidth, width=0.47\textwidth}
\vspace{-3mm}
\caption{\small The solution of isothermal sphere in $u-v$ plane (thick solid line). As the curve
spirals into the point $(u,v)=(1,2)$, the density contrast between the center and the boundary
increases and the sphere becomes singular isothermal sphere $\rho \sim \frac{1}{r^2}$.
Two thin solid lines correspond to the energy and the temperature bound respectively.
All equilibrium isothermal sphere with $RE/GM^2 = \lambda$ surrounded by non-conducting wall
must be located at the intersection between the spiral curve and a straight line with slope $\lambda^{-1}$.
$GM\beta/R = \eta$ for all equilibrium isothermal sphere surrounded by conducting wall has to intersect
with the spiral curve.}
\label{fig1}
\end{figure}

If we solve this equation numerically using the boundary conditions of $v=0$ at $u=3$ and $\frac{dv}{du}=-5/3$ at $(u,v)=(3,0)$ 
corresponding to $\psi(0) = \psi'(0)=0$,
we obtain a spiraling curve on $u-v$ plane as shown in \fig~\ref{fig1} (also see Fig. 2 in \citet{padma1989}).
The equilibrium isothermal sphere must exist on this curve in $u-v$ plane.
Since the enclosed mass within $r$ of singular isothermal sphere diverges as $r$ increases, it is more physically meaningful to 
consider the cut-off at radius $R$. Two simple boundary conditions have been considered: non-conducting spherical wall where no energy 
is transfered and conducting spherical wall with fixed temperature $T$, where energy is transfered.

For non-conducting wall, the energy of self-gravitating system is conserved.
Using \eqs~\ref{vvar} and \ref{uvar},
the dimensionless energy $\lambda$ is defined as
\begin{equation}
\lambda  \equiv  \frac{RE}{GM^2} = \frac{1}{v_0} \big( u_0 - \frac{3}{2} \big)
\label{lambda}
\end{equation} 
where $E$ is total energy and $M$ is total enclosed mass within $R$. Subscript 0 indicates the value at $\xi=R/L_0$.
If we rewrite this equation in slightly different form, we have a linear line on $u-v$ plane with the slope given by $\frac{1}{\lambda}$
\begin{equation}
v_0 = \frac{1}{\lambda} \big( u_0 - \frac{3}{2} \big).
\label{uvline}
\end{equation}
In order for the system surrounded by non-conducting wall to be in isothermal equilibrium 
$\lambda$ must be greater than $\lambda_c=-0.335$ \citep{antonov1962,bt2008,padma1989,padma1990}.  

For conducting wall, the temperature of the system is conserved.
The dimensionless inverse temperature $\eta$ is defined as
\begin{equation}
\eta  \equiv  \frac{GM \beta}{R} = v_0.
\label{eta}
\end{equation}
Therefore, as shown in \fig~\ref{fig1}, $\eta$ must be smaller 
than $\eta_c=2.52$ for this system to be in isothermal equilibrium \citep{bt2008,chavanis2002a}.

If we introduce the concept of temperature $T$ of a stellar system with $N$ stars,
the heat capacity of the system is
\begin{equation}
C \equiv \frac{dE}{dT}= - \frac{3}{2}N k_B < 0 .
\end{equation}
We see that the heat capacity is negative: the more energy the system loses, the hotter the system
becomes. This apparently paradoxical phenomenon is seen not only in stellar system but also in
any finite system governed by gravity \citep{bt2008}.
The heat capacity $C$ can also be written using $\lambda$ and $\eta$ \citep{chavanis2002a}
\begin{equation}
C \equiv \frac{dE}{dT}= \frac{dE}{d\beta} \frac{d\beta}{dT}
= \frac{Nk_B}{M}\beta^2 \frac{dE}{d\beta} = -Nk_B \eta^2 \frac{d\lambda}{d\eta} .
\end{equation}
Using \eqs~\ref{uveq1},\ref{uveq2},\ref{lambda} and \ref{eta}, we get
\begin{eqnarray}
\frac{d\lambda}{d\xi} &=& \frac{1}{2v\xi}(4u^2 + 2uv - 11u +3)\\
\frac{d\eta}{d\xi} &=& \frac{v}{\xi} (u-1).
\end{eqnarray}
Thus
\begin{equation}
C = - Nk_B \frac{4u^2 + 2uv - 11u + 3}{2(u-1)}.
\label{heatcapacity}
\end{equation}
Using $\lambda$ and $\eta$, we can regard the stellar
system as a thermodynamical system with total energy $E$ and temperature $T$.

\begin{figure}
\center
\epsfig{figure=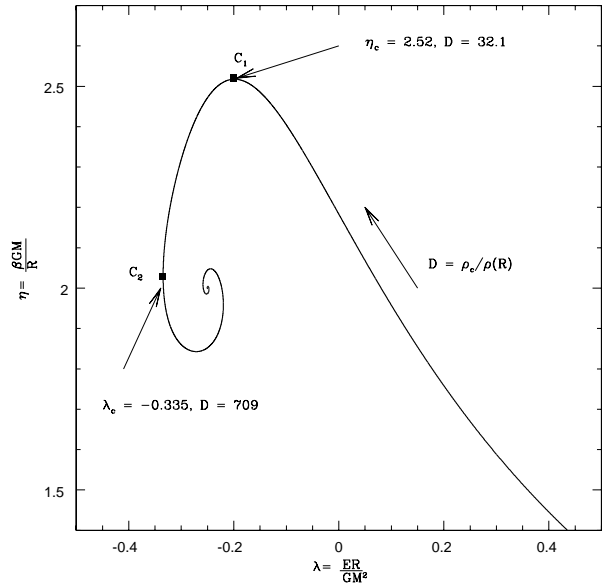, height=0.47\textwidth, width=0.47\textwidth}
\vspace{-3mm}
\caption{
The $\eta(\lambda)$ curve for isothermal sphere. The heat capacity is positive until the system
reaches the point $\mbox{C}_1$ and becomes negative after $\mbox{C}_1$.
After $\mbox{C}_1$, the isothermal sphere surrounded by a conducting wall is unstable.
After $\mbox{C}_2$, the isothermal sphere surrounded by a non-conducting wall is unstable.}
\label{fig2}
\end{figure}

We show $\eta$ as a function of $\lambda$ in \fig~\ref{fig2} (also see Figure 7.1 in \citet{bt2008}).
As the isothermal sphere follows the solid spiral curve from the lower right corner, the density contrast between
the center and the boundary $D=\frac{\rho_c}{\rho(R)}$ increases and the heat capacity characterized by $\frac{d\lambda}{d\eta}$ 
varies over the range between $-\infty$ to $+\infty$. 
If the isothermal sphere surrounded by a non-conducting wall passes the point $\mbox{C}_2$ where $D=709$,
the system is unstable although the heat capacity is positive
\citep{antonov1962,bt2008,chavanis2002a,hk1978,lw1968,katz1978,padma1989,padma1990}.
This instability originally introduced by \citet{antonov1962} is later called the gravothermal catastrophe \citep{lw1968}.
If the isothermal sphere surrounded by a conducting wall with fixed temperature 
$T$ passes the point $\mbox{C}_1$ where $D=32.1$, the system is unstable \citep{bt2008,chavanis2002a,hk1978,katz1978}. 
However it does not lead to `core-halo' structure \citep{chavanis2002a} contrary to 
the case of non-conducting wall \citep{padma1990}.

\begin{figure}
\center
\epsfig{figure=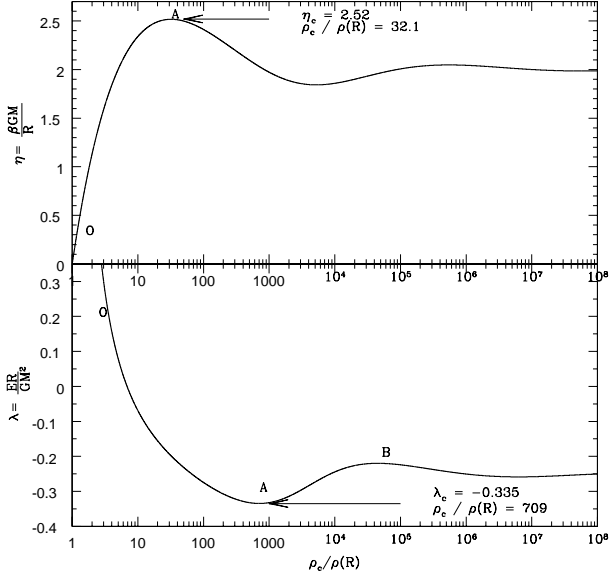, height=0.47\textwidth, width=0.47\textwidth}
\vspace{-3mm}
\caption{
Upper panel: $\eta$ as a function of the density contrast $D=\frac{\rho_c}{\rho(R)}$
for an isothermal sphere surrounded by a conducting wall.
All points beyond the first position where the $\frac{d\eta}{dD} = 0$
are known to be unstable \citep{chavanis2002a}.
Lower panel: $\lambda$ as a function of density contrast $D=\frac{\rho_c}{\rho(R)}$
for an isothermal sphere surrounded by a non-conducting wall.
All points beyond the first position where $\frac{d\lambda}{dD} = 0$
are known to be unstable \citep{katz1978}.
}
\label{fig3}
\end{figure}

In \fig~\ref{fig3} we show the $\lambda$ and $\eta$ as a function of $D$.
Making Boltzmann-Gibbs entropy extreme (i.e. $\delta S = 0$) in the microcanonical ensemble, the equilibrium solution of self-gravitating system
is isothermal. At every point on $\lambda(D)$ curve in the lower panel of \fig~\ref{fig3}, 
$\delta S = 0$ and at critical points where $\frac{d\lambda}{dD} = 0$, $\delta^2 S = 0$ \citep{lw1968,padma1989,padma1990}.
\citet{padma1989,padma1990} showed that the entropy of the system is in a local maximum at any point on the branch 
OA (i.e. $\delta^2 S < 0$) and the entropy of the system is in a local minimum along the branch AB 
(i.e. $\delta^2 S > 0$), except at A which is a saddle point.
The branch OA is stable and the branch AB is unstable \citep{padma1989}.
It is known that all points beyond A are unstable based on previous analyses \citep{katz1978,katz1979}.
Similarly \citet{chavanis2002a} showed that using the free energy ($F=E-TS$) instead of entropy $S$, the system in the canonical 
ensemble (conducting wall) becomes unstable after the first position (i.e. marked by A) where $\frac{d\eta}{dD} = 0$ in the upper panel of \fig~\ref{fig3}.

\section{Self-gravitating isothermal sphere embedded in potential well}
The boundary condition surrounding isothermal sphere in previous works has been either conducting or non-conducting wall. 
These boundary conditions were necessary to make the problem simple. 
Here we replace the boundary conditions with the realistic potential function and demonstrate that the potential well
is similar to the heat bath if the potential well is deep compared with the potential depth of central embedded system. 

In this work, a spherical stellar system is considered to be embedded in potential well whose
center coincides that of the stellar system under consideration. 
Then we can follow the same procedure of Section 2 by adding extra terms. 
If we include the external potential well $\phi_{ext}$, we can write the potential and kinetic energy of the system as follows
\begin{eqnarray}
U & = & -\int^R_0 \frac{GM(r)}{r} \frac{dM}{dr} dr + \int^R_0 \frac{dM}{dr} \phi_{ext} \label{extpotential}\\
K & = & \frac{3}{2} \frac{M}{\beta} = \frac{GM_0^2}{L_0} \frac{3}{2}
\int^{\xi_0}_0 \frac{dm}{d\xi} d\xi = \frac{GM_0^2}{L_0} \frac{3}{2}
\int^{\xi_0}_0 n\xi^2 d\xi , \nonumber \\
\end{eqnarray}
where $R$ is the size of embedded system and $\xi_0=R/L_0$. 
We use Plummer potential as $\phi_{ext}$ and approximate it as a harmonic potential 
near the center by Taylor expansion:
\begin{equation}
\phi_{ext}= - \frac{GM_e}{a}\frac{1}{\sqrt{1+\frac{r^2}{a^2}}}
\sim - \frac{GM_e}{a}(1-\frac{r^2}{2a^2}).
\end{equation}
where $M_e$ and $a$ are the mass and the scale length of Plummer potential.
Here, the harmonic approximation is valid as long as the embedding potential has 
the spatial scale much larger than the scale of the stellar system, as is often the case
with the stellar systems in Galactic center (i.e. $\le 0.4$pc core radius, see e.g. \citet{eckart2005,figer1999}) 
embedded in much larger bulge (i.e. $\approx 0.56$ Kpc characteristic scale length modeled by Dehnen profile \citep{dehnen1993}, 
see e.g. \citet{bm1999}), or the galaxies in cluster (i.e. radial distribution of galaxies modeled by 
broken power-law has a scale radius $20\%$ smaller than the cluster radius, see e.g. \citet{marel2000,mo2010}).

\begin{figure}
\center
\epsfig{figure=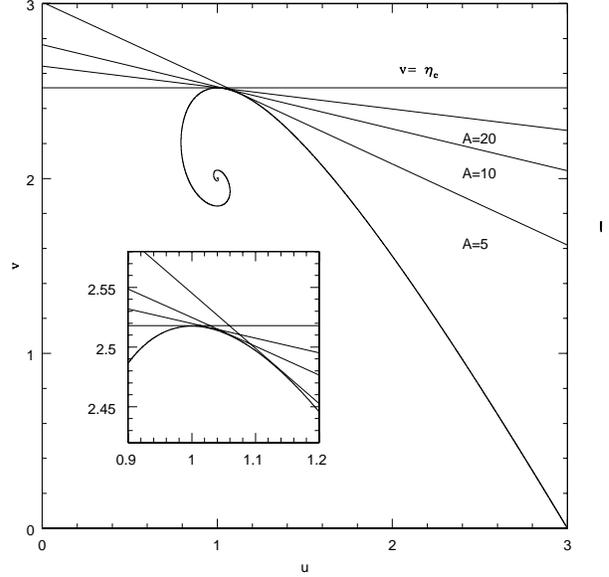, height=0.47\textwidth, width=0.47\textwidth}
\vspace{-3mm}
\caption{\small
The solution of isothermal sphere in $u-v$ plane (thick solid line) and lines
determined by \eq~\ref{uvline3} for given $A$.
Three thin solid lines have three different slopes determined by $\lambda_c$ when considering
the external potential wells. If the value $A$ is large, the ratio between the depth of
external potential and potential of central embedded system is also large.
As the external potential depth becomes deep (large $A$), the line associated with the energy bound is close to
the straight line $v=\eta_c$, which is temperature bound for isothermal sphere surrounded by heat bath.
The inset figure magnifies the region of $u-v$ space over the domain $[0.9,1.2]\otimes[2.42,2.58]$ and 
shows that as $A$ increases, the point where the tangent line and the curve meet becomes close to the point 
$(u,v)=(1.0,2.52)$ where the straight line $v=\eta_c$ meets the curve.}
\label{fig4}
\end{figure}

Then we can rewrite \eq~\ref{extpotential} using \eqs~\ref{scalevar},\ref{dimlessvar} and \ref{dimlessvar2} as
\begin{eqnarray}
U & = & -\frac{GM_0^2}{L_0} \int^{\xi_0}_0 mn\xi d\xi \nonumber \\
&-& \frac{GM_0 M_e}{a} \int^{\xi_0}_0 n\xi^2\left(1-\frac{L_0^2}{2a^2}\xi^2\right)
d\xi.
\end{eqnarray}
Then the total energy $E$ is

\begin{eqnarray}
E \equiv U + K & = & \frac{GM_0^2}{L_0} \int^{\xi_0}_0 \left(\frac{3}{2}n \xi^2 - mn\xi\right) d\xi\nonumber \\
&  -& \frac{GM_0 M_e}{a} \int^{\xi_0}_0 n\xi^2 \left( 1 - 
\frac{L_0^2}{2a^2}\xi^2\right) d\xi \nonumber\\
 & = & \frac{GM_0^2}{2L_0} \int^{\xi_0}_0 \Big[ 3n\xi^2 - 2mn\xi - 
2\left(\frac{L_0}{a}\frac{M_e}{M_0}\right) n\xi^2 \nonumber\\
& + &    \left(\frac{M_e}{M_0} \frac{L_0^3}{a^3}\right) 
n\xi^4 \Big] d\xi \nonumber \\
\end{eqnarray}
By defining the mean density within $r$ as $\bar{\rho}_r\equiv\frac{3M(r)}{4\pi r^3}$,
$\frac{m}{\xi^3}=\frac{M(r)}{r^3}\frac{L_0^3}{M_0}=\frac{1}{3}\frac{\bar{\rho}_r}{\rho_c}=\frac{1}{3}\bar{n}$.
We can write the dimensionless energy $\lambda$ of the system as follows.
\begin{equation}
\lambda  \equiv  \frac{RE}{GM^2} = \frac{\xi_0}{2m_0^2}(2n_0\xi_0^3 - 3m_0 - 2Am_0
+ Bm_0\xi^2_0 -\frac{2B\bar{n}}{15}\frac{\xi^6_0}{m_0^2})\nonumber \\
\end{equation}
where $A \equiv \frac{L_0}{a}\frac{M_e}{M_0}, B \equiv \frac{L_0^3}{a^3}\frac{M_e}{M_0},
n_0=n(\xi_0),m_0=m(\xi_0)$.
Using $u$ and $v$, this can be rewritten as
\begin{equation}
v_0  = \frac{u_0 - (1.5+A)}{\lambda - \frac{1}{6}\frac{\rho_e(0)}{\rho(R)}
u_0 + \frac{1}{15}\frac{\rho_e(0)}{\rho(R)}\frac{\bar{\rho}_R}{\rho(R)} u_0^2}
\label{newuvline}
\end{equation}
where $\rho_e(0) = \frac{3M_e}{4\pi a^3}$.
Recalling that $u = \frac{\xi e^{\psi}}{\psi'} = \frac{n\xi^3}{m}$, $\frac{m}{\xi^3}=\frac{1}{3}\bar{n}$, 
$\frac{\bar{\rho}_R}{\rho(R)} = \frac{3}{u_0}$,
we can rewrite \eq~\ref{newuvline} as
\begin{equation}
v_0  = \frac{u_0 - (1.5+A)}{\lambda - \frac{1}{10}\frac{\rho_e(0)}{\bar{\rho}_R}}.
\label{uvline2}
\end{equation}
This new relation gives a linear line which is different from \eq~\ref{uvline}.
In \eq~\ref{uvline2}, the term $A \equiv \frac{L_0}{a}\frac{M_e}{M_0}$ is the ratio of depth of 
the external potential well and isothermal sphere at the center. Also the previous slope determined as $\lambda^{-1}$ 
is modified to $\left(\lambda-\frac{1}{10}\frac{\rho_e(0)}{\bar{\rho}_R}\right)^{-1}$ for a given mean density of 
the embedded isothermal sphere $\bar{\rho}_R$ and central density of the external potential $\rho_e(0)$.
Since we are interested in the case where the potential well is deep and not much affected
by the evolution of the central isothermal sphere, 
$A \gg 1$ and the slope $\left(\lambda-\frac{1}{10}\frac{\rho_e(0)}{\bar{\rho}_R}\right)^{-1}$ should be small 
(or $\lambda-\frac{1}{10}\frac{\rho_e(0)}{\bar{\rho}_R}$ should be large) but negative
in order for the line to intersect with $u-v$ curve.
On the other hand, $\frac{\rho_e(0)}{\bar{\rho}_R}=\frac{M_e}{M_0}\frac{R^3}{a^3} \ll 1$ if the size of the sphere $R$ 
is assumed to be much smaller than the characteristic scale $a$ of external potential 
and as a result, the term $\frac{M_e}{M_0} (\frac{R}{a})^3$ is significantly smaller than 1, which is appropriate assumption.
Therefore $\lambda-\frac{1}{10}\frac{\rho_e(0)}{\bar{\rho}_R} \sim \lambda$ and \eq~\ref{uvline2} approximately becomes
\begin{equation}
v_0  = \frac{1}{\lambda}\{u_0 - (\frac{3}{2}+A)\}.
\label{uvline3}
\end{equation}
For $A \gg 1$, we can determine maximum $\lambda$, which turns out to be the tangent line to the spiral curve on $u-v$ plane
in \fig~\ref{fig2}. \fig~\ref{fig4} shows the same spiral curve as \fig~\ref{fig2} with several lines determined 
by \eq~\ref{uvline3} with different values of $A$.

For each line, there is an associated critical $\lambda_c$. 
However the $\lambda_c$ is now upper boundary in contrast to the previous case of isothermal sphere surrounded by
non-conducting wall, which gives lower boundary $\lambda_c=-0.335$.
If $\lambda$ is larger than $\lambda_c$, the self-gravitating isothermal sphere 
surrounded by the potential well can not be in equilibrium.
As the potential depth of surrounding potential well becomes deeper (larger $A$), the tangential line becomes close to the
line $\eta_c=2.52$, which sets the minimum temperature of equilibrium isothermal sphere surrounded by 
thermally conducting wall. In other words, when the potential well is very deep compared to the potential 
depth of central self-gravitating system, the potential well behaves like a heat bath, which can heat up the embedded system.

\begin{figure}
\center
\epsfig{figure=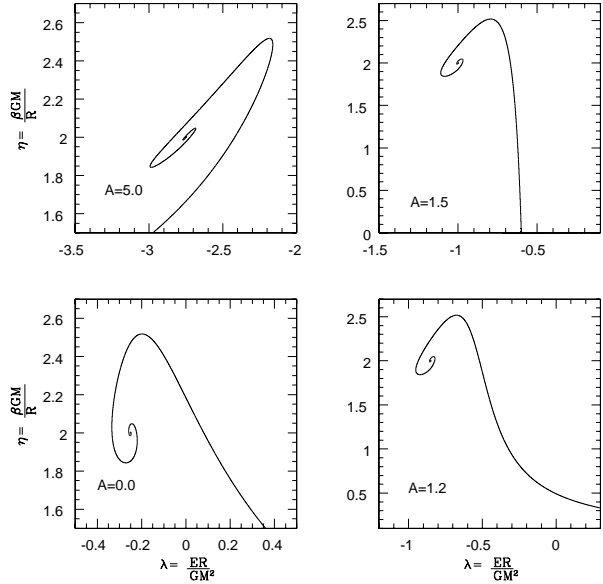, height=0.470\textwidth, width=0.47\textwidth}
\vspace{-3mm}
\caption{\small The $\eta(\lambda)$ curve for the isothermal sphere embedded in
the external potential well. If $A < 1.5$, the heat capacity is
initially positive and grows to infinity $\frac{d\eta}{d\lambda}=0$, then becomes negative. However
if $A > 1.5$, heat capacity is initially negative and increases to positive value, then becomes
infinity $\frac{d\eta}{d\lambda}=0$ changing its sign. And if $A = 1.5$, 
$C$ is $0$ when the system is homogeneous (i.e. density contrast is 1), then grows to infinity as the system goes inhomogeneous 
density structure (i.e. increasing density contrast).}
\label{fig5}
\end{figure}

The heat capacity of the system surrounded by the potential well can also be written as
\begin{equation}
C = - Nk_B \frac{4u^2 + 2uv - (11+2A)u + 3+2A}{2(u-1)}
\end{equation}
in the same way to obtain \eq~\ref{heatcapacity}.
We can get the relation between $\lambda$ and $\eta$ with different $A$ as shown in \fig~\ref{fig5}.
Please note that lower left panel of \fig~\ref{fig5} is the same as \fig~\ref{fig2} (i.e. $A=0.0$).
There are critical points where $C=0$ or $C \rightarrow \infty$ in \fig~\ref{fig5}. When there is no
potential well ($A=0.0$) or shallow potential well ($A=1.2$), a critical point appears first at $C \rightarrow \infty$.
If potential well becomes deep and $A$ finally exceeds 1.5,
a critical point first appears when $C=0$. As the potential well becomes even deeper,
the location where $C=0$ in $\lambda-\eta$ plane is close to the location 
where $C \rightarrow \infty$. This means that the instability
of the self-gravitating system surrounded by the potential well appears on the nearly same point
where the instability of the system surrounded by heat bath occurs. 
As also seen in \fig~\ref{fig4}, if $A$ is larger or smaller than 1.5,
there is upper or lower bound value of $\lambda_c$ for equilibrium.

\begin{figure}
\center
\epsfig{figure=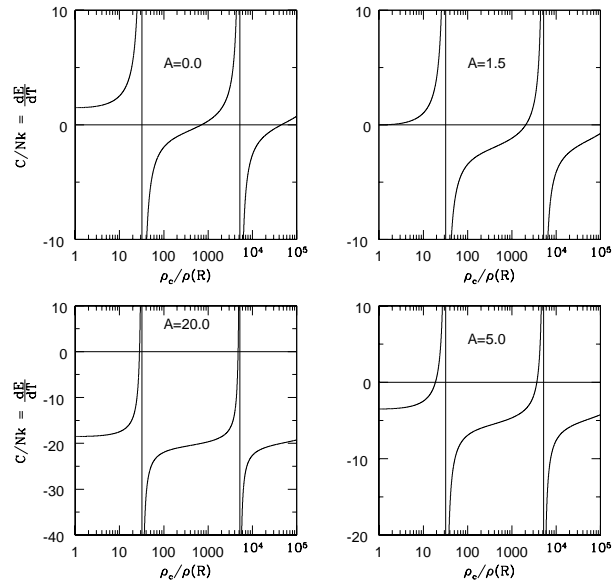, height=0.47\textwidth, width=0.47\textwidth}
\vspace{-3mm}
\caption{Heat capacity as a function of density contrast $D\equiv\frac{\rho_c}{\rho(R)}$. The heat capacities
of isothermal spheres with potential wells diverge at the same point $D=32.1$.
However the points with $C=0$ below which the heat capacity is negative and gravothermal catastrophe occurs
are different.
As the potential well becomes deep (i.e. $A\gg1.5$), the value of $D$ where $C = 0$, is close to
the value of $D$ at $C \rightarrow \infty$.}
\label{fig6}
\end{figure}

The heat capacity of the system is shown in \fig~\ref{fig6} for different $A$. 
Without external potential well ($A=0$), the heat capacity becomes negative after the density contrast $D$ is greater
than 32.1. Then the negative heat capacity makes the system gravitationally unstable and collapse.
When $D>709$, the heat capacity becomes positive but the system is known to be unstable \citep{katz1978,katz1979}.
When the external potential becomes deep (e.g. $A=20$ in \fig~\ref{fig6}), 
the heat capacity of the system is negative even if it has homogeneous matter distribution with small $D$. 
However in this case, the gravothermal collapse is restrained since the system is embedded in deep potential well, 
which behaves like a heat bath with high temperature and heats up the system.
Then the heat capacity becomes positive when $D$ is very close to 32.1, which is the critical density contrast where
the isothermal sphere surrounded by the heat bath becomes unstable.
This also supports the argument that the external potential behaves effectively like a heat bath.

\begin{figure}
\center
\epsfig{figure=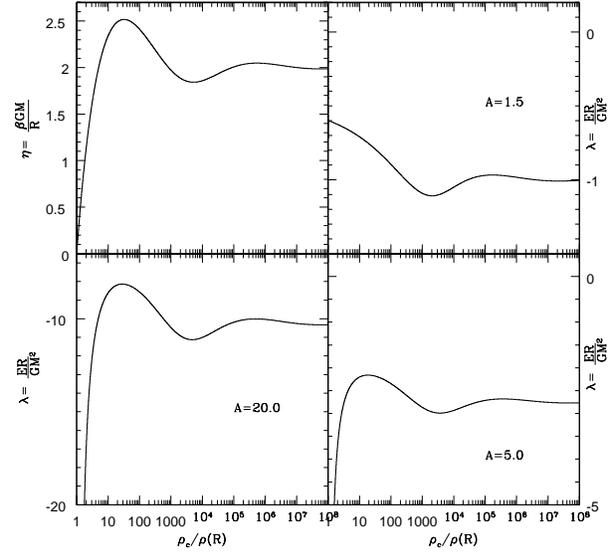, height=0.45\textwidth, width=0.45\textwidth}
\vspace{-3mm}
\caption{Upper left panel shows $\eta$ as a function of density contrast $D$ which is the same as the upper panel of \fig~\ref{fig3}.
Other panels show $\lambda$ with different $A$, as a function of density contrast $D$. Please note that 
the lower panel of \fig~\ref{fig3} is the case where $A=0$.
Upper right panel ($A=1.5$) shows that, in contrast to the case of $A=0.0$, the point where $\frac{d\lambda}{dD} = 0.0$ first occurs 
at $D=1.0$ and then appears again at $D\sim2000$.
Lower right panel ($A=5.0$) shows that $\lambda$ reaches a upper bound beyond which the isothermal equilibrium state does not exist, 
at $D \sim 20.0$ where $\frac{d\lambda}{dD} = 0.0$.
Lower left panel ($A=20$) shows that the upper bound $\lambda$ appears at $D \sim 29.0$, which is close to $D=32.1$
where the instability of self-gravitating isothermal sphere enclosed by heat bath occurs. If $A$ keeps increasing, the point where 
the maximum $\lambda$ appears becomes close to $D=32.1$.
}
\label{fig7}
\end{figure}

In \fig~\ref{fig7}, we show the $\lambda(D)$ curves of the embedded isothermal sphere with different $A$ 
(i.e. different potential depth) and compare them to the $\eta(D)$ curve of the isothermal sphere.
The upper left panel shows the $\eta$ as a function of $D$ which is the same as the upper panel of \fig~\ref{fig3}.
Other panels show how the $\lambda(D)$ of isothermal sphere embedded in potential well changes with different potential depth.
When $A < 1.5$, there is a lower bound $\lambda$ for equilibrium, however, if $A > 1.5$, there is
a upper bound for equilibrium. As $A$ becomes large,
the density contrast at upper bound $\lambda$ approaches to 32.1, beyond which the self-gravitating isothermal sphere within 
conducting wall, becomes unstable.  
This indicates that as potential well becomes deep, it approaches to the same boundary condition as the conducting wall (i.e. heat bath).

\section{Dynamical evolution of stellar system within potential well}
As shown in Section 3, the external potential well, if it is deep enough compared with the self-gravitating system in 
the center, increases the temperature of self-gravitating system and becomes effectively the same boundary condition as conducting wall.
Therefore it is interesting to study how self-gravitating system in potential well evolves.
We use GPU version of the direct $N$-body simulation code NBODY6 \citep{aarseth1999} and simulate the dynamical evolution of 
stellar system embedded in the external potential well which is assumed to be fixed during the evolution. 

We generate four simple models. In each model, a stellar system following Plummer density profile 
is enclosed by the external potential wells with different depths. 
The central stellar system is generated from Plummer initial condition built in NBODY6, using 10000 particles,
and has the following form of potential-density pair.
\begin{eqnarray}
\phi = - \frac{GM}{a} \Big[ 1 + (\frac{r}{a})^2 \Big]^{-1/2}\\
\rho = \frac{3M}{4\pi a^3} \Big[ 1 + (\frac{r}{a})^2 \Big]^{-5/2}
\end{eqnarray}
The Hermite scheme in NBODY6 requires force vectors to be differentiated
up to the third order \citep{aarseth1985,aarseth2003}. 
The external Plummer potential ($\phi_e$) is already implemented in NBODY6.
The external Plummer potential has the following potential-density pair with different mass $M_e$ and scale length $R$.
\begin{eqnarray}
\phi_e = - \frac{GM_e}{R} \Big[ 1 + (\frac{r}{R})^2 \Big]^{-1/2}\\
\rho_e = \frac{3M_e}{4\pi R^3} \Big[ 1 + (\frac{r}{R})^2 \Big]^{-5/2}
\label{extpot}
\end{eqnarray}
Please note that the notation $a$ and $R$ used in this section are different from those in Section 3.

\begin{table}
\begin{center}
\caption{Properties of initial models\label{tab1}}
\begin{tabular}{ccccccc}\\
\hline\hline
Model & $N$ & $M_e$ & $R^2$ & $E_{tot}$\tablenotemark{\dagger} & $K$ & $\frac{M_e}{R}\frac{a}{M}$\\
\hline
Model1 & 10000  &    100.0   &  10    &  -28.6 & 1.666 & 18.62 \\
Model2 & 10000  &    100.0   &  50    &  -14.0 & 0.450 & 8.33 \\
Model3 & 10000  &    100.0   &  75    &  -11.6 & 0.369 & 6.80 \\
Model4 & 10000  &    100.0   &  100   &  -10.1 & 0.333 & 5.89 \\
\hline
\end{tabular}
\tablenotetext{\dagger}{$E_{tot} = W_s + W_e + K$}
\end{center}
\end{table}


We use the following general units used in $N$-body simulation:
\begin{equation}
time: \frac{GM^{5/2}}{(-4E)^{3/2}} \quad\quad length: \frac{GM^2}{-4E} \quad\quad mass: M
\end {equation}
where $G=1$,$M=1$ and $E=-\frac{1}{4}$ \citep{hm1986}.
Using these units, $a=3\pi/16$ \citep{ahw1974,hh2003}.
Initial conditions of our models are listed in \tab~\ref{tab1}. Masses for external potential well are all set to 100 in $N$-body 
simulation unit. Square of scale length $R^2$ is different and shown in $N$-body simulation unit.

\begin{figure}
\center
\epsfig{figure=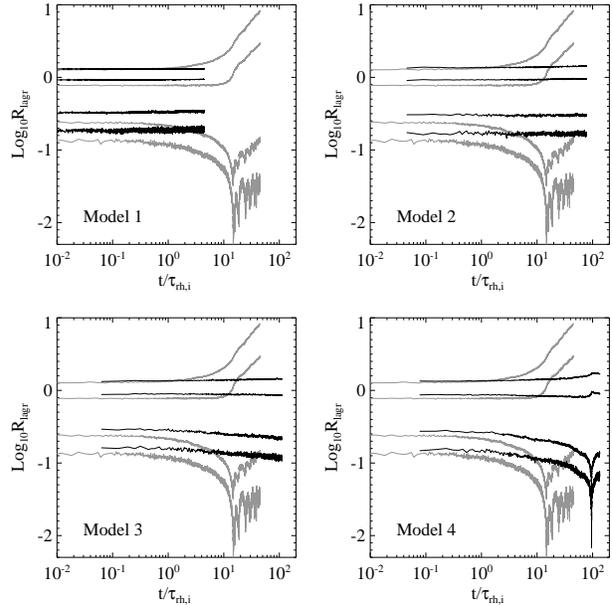, height=0.47\textwidth, width=0.47\textwidth}
\vspace{-3mm}
\caption{The comparison between the evolution of Lagrangian radii of each model
and isolated Plummer model (black line).
Clearly we see that core collapse time of the system embedded in
the potential well is longer than that of isolated Plummer model.
When the external potential depth is deep, this discrepancy increases.
}
\label{fig8}
\end{figure}

In \fig~\ref{fig8}, we show the evolution of Lagrangian radii which contain 1,5,50 and 85\%
of the total mass of embedded stellar system.
The black line in each figure is the result of isolated Plummer model and shown for comparison.
Color lines are the result from 4 different models.
Time is scaled by initial half mass relaxation time \citep{spitzer1987}.
\begin{equation}
T_{rh,i} =  \frac{<v^2>^{3/2}}{15.4 G^2 m \rho \ln \Lambda}
\end{equation}
Using $\rho \sim \frac{3M}{8\pi R_h^3}$ and $N=\frac{M}{m}$, $T_{rh,i}$ used in NBODY6 can be rewritten as
\begin{equation}
T_{rh,i} =  \frac{8\pi}{3} \frac{N <v^2>^{3/2} R_h^3}{15.4 G^2 M^2 \ln \Lambda}
\label{rlxtime}
\end{equation}
where $\ln \Lambda=\ln(\gamma N)$ is Coulomb logarithm determined by two body relaxation and $\gamma$ is
usually 0.4 \citep{aarseth2003}.
For the isolated Plummer model, the inner Lagrangian radii (1,5\%) decrease and the outer Lagrangian radii (50,85\%)
increase due to the gravothermal catastrophe which leads to the core collapse occurring at ($15\sim16 T_{rh,i}$) 
as seen in \fig~\ref{fig8}. Then later the inner Lagrangian radii increase due to the outward heat flow by the two body interaction.

However the evolution of Lagrangian radii of the stellar system is retarded if it is surrounded by the external 
potential well, as shown with color in each panel of \fig~\ref{fig8}. The potential depth ratio between 
the external potential and the embedded stellar system is $\frac{M_e}{R}\frac{a}{M}$ and 
increases from 5.89 for Model4 to 18.62 for Model1, as seen in \tab~\ref{tab1}.
NBODY6 time step is scaled by $T_{rh,i}$ for each model using \eq~\ref{rlxtime}. 
As the potential well becomes deeper, core collapse time becomes longer than that of isolated system or
core collapse does not occur.

Another interesting point is the concentration of the embedded stellar system compared to the isolated system. 
Since the potential well behaves like a heat bath and increases the velocity dispersion of the embedded stellar system,
the central region of the embedded stellar system expands due to the increased velocity dispersion while the radius of outer boundary 
is fixed because the system is confined by the external potential well. In the case of deep potential well 
(i.e. Model1 in lower left panel of \fig~\ref{fig8}), it is clearly seen that inner Lagrangian radii of embedded system 
are respectively larger than those of isolated system as seen in \fig~\ref{fig8}.

\begin{figure}
\center
\epsfig{figure=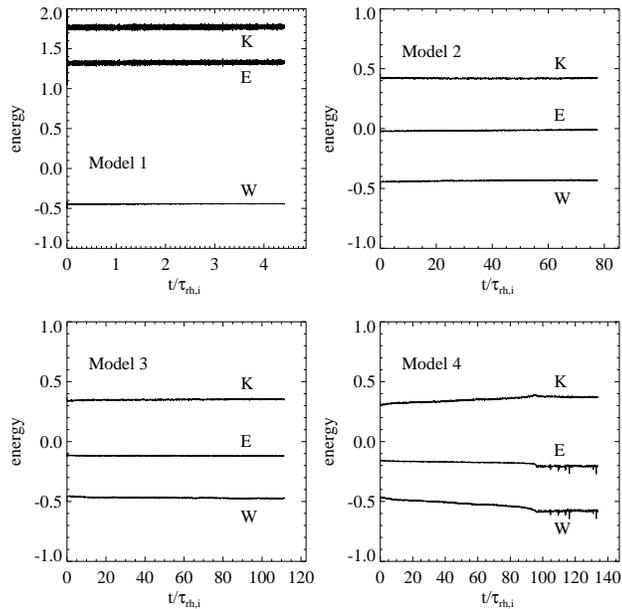, height=0.47\textwidth, width=0.47\textwidth}
\vspace{-3mm}
\caption{\small The evolution of kinetic, potential and
total energies of embedded stellar system for each model.
Here the external potential energy is not included in the potential energy.
It is easily seen that Model1 is gravitationally unbound. Other models are bound by gravity and
have negative heat capacity, thus expect to have core collapse.
However the external potential increases the velocity dispersion of embedded system, which makes
the relaxation time longer than that of isolated Plummer model.
Also, after a long time, the models suppose to reach the thermal equilibrium due to
the interaction with the external potential well that behaves like a heat bath.
}
\label{fig9}
\end{figure}

We also show the evolution of kinetic ($K$), potential ($W$, without the external potential) and total 
energy ($E$) of embedded stellar system in \fig~\ref{fig9}. 
$N$-body simulation time is scaled using \eq~\ref{rlxtime}. 
The initial $E$, $K$ and $W$ of isolated system are $-\frac{1}{4}$,
$\frac{1}{4}$ and $-\frac{1}{2}$ respectively.
For Model4, we see that the initial $E$ is larger than $-\frac{1}{2}$, but less than 0. 
Thus it is gravitationally bound and gravothermal catastrophe occurs due to the negative heat capacity.
The initial $K$ and $W$ are about $0.30$ and $-0.47$ respectively. If we compare the initial potential and kinetic energies
of the Model4 to those of isolated model, we see that the kinetic energy significantly increases from 0.25 to 0.30.
The total energy of the Model3 and Model2 are also negative, but more close to 0 (i.e. less tightly bounded than
Model4). And the difference of kinetic energy of our models from that of isolated Plummer model is 
larger than the difference of potential energy. 
Model1 is gravitationally unbound (i.e. $E$ is positive). Kinetic energy of the Model1 is much larger than
the values of other models. As the potential depth becomes deeper,
the kinetic energy increases due to the heating by the potential well.
Thus velocity dispersion of embedded system increases and this leads to large $T_{rh,i}$ (see \eq~\ref{rlxtime}).
This means that the potential well makes the relaxation process slow.
While the embedded stellar system heated by the potential well tends to expand due to increased velocity dispersion,
the outer parts can not expand because the potential well confines the system. 

\begin{figure}
\center
\epsfig{figure=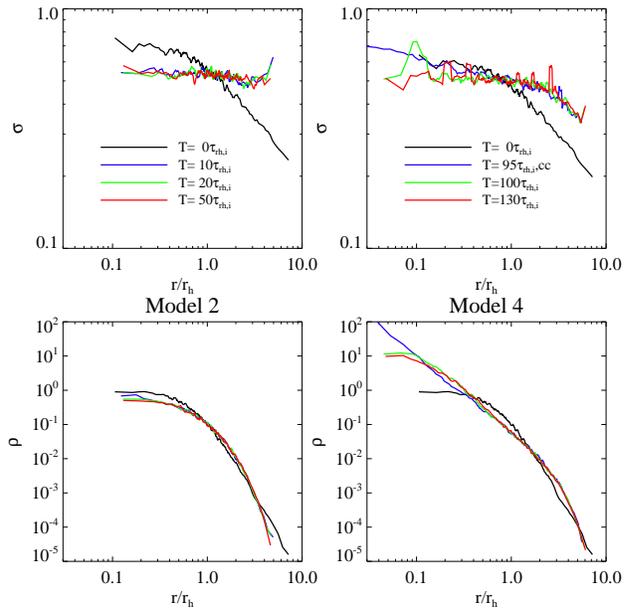, height=0.47\textwidth, width=0.47\textwidth}
\vspace{-3mm}
\caption{\small The snapshots of radial profile of density and velocity dispersion of Model2 and Model4.
Model2 quickly reaches the thermal equilibrium with the external potential well and thus its velocity dispersion becomes
isothermal. The final radial density profile of Model2 in the lower left panel is slightly changed from the initial density profile.
Model4 experiences core collapse at $t=30T_{rh,i}$. Although the gradient of velocity dispersion
is reduced with time, there is still a gradient observed in outer region ($r > r_h$) of the Model4
after core collapse. The radial density increases by two orders of magnitude at core collapse and
by several tens at final quasi-equilibrium state.
}
\label{fig10}
\end{figure}

In \fig~\ref{fig10}, we show the snapshots of density and velocity dispersion profiles for two models (Model2 and Model4).
Radius is scaled by initial half mass radius.
In the left panels (Model2), black line corresponds to the initial density (lower panel) and 
velocity dispersion (upper panel) profiles.
The blue, green and red lines represent the profiles at $T=10, 20~\mbox{and}~50 T_{rh,i}$ respectively.
In the right panels (Model4), the black, blue, green and red lines are the profiles at $T=0, 95, 100~\mbox{and}~130 T_{rh,i}$ respectively. 
Note that, for Model4, core collapse occurs at $T=95T_{rh,i}$.

For Model2, we see that the velocity dispersion profile becomes isothermal after 
a few times of $T_{rh,i}$. Core-collapse does not occur even if $T=50T_{rh,i}$, and the density profile varies 
slightly from the initial profile: central concentration decreases slightly and outer region is more sharply truncated.
Model2 is less gravitationally bound system than Model4 (see \fig~\ref{fig9}). 
Initially there is a gradient of velocity dispersion (i.e. the center is warmer than the outer). 
However, as the stellar system embedded in a deep potential well dynamically evolves, 
the velocity dispersion becomes isothermal as a result of thermal equilibrium with the potential well 
(effectively a heat bath) although the slight gradient is seen beyond $R_h$.
From the density profile in lower left panel of \fig~\ref{fig10} we can see that 
the embedded stellar system becomes less concentrated because
the inner region heated by the potential tends to expand and outer region confined by the potential
can not.

On the other hand, we observe core collapse and gradient of velocity dispersion profile for the Model4 
in the right panels of \fig~\ref{fig10}. As seen in upper left panel of \fig~\ref{fig8}, core collapse occurs at $95T_{rh,i}$ and the 
density profile (blue) at core collapse has large concentration and shows `core-halo' structure. 
Although Model4 is surrounded by external potential well, it is gravitationally bound system with 
negative energy (see upper left panel of \fig~\ref{fig9}).
Therefore the self-gravity of the embedded system dominates its evolution.
The velocity dispersion was not isothermal initially and the gradient of velocity dispersion still exists after core collapse,
in contrast to the case of Model2 where the initial gradient of velocity dispersion is ironed out fast due to the heating 
by deep external potential well.

From these $N$-body simulation results we see that 
the external potential well makes the relaxation process of embedded self-gravitating system slow by heating 
the system and retards core collapse, or prohibits core collapse if the potential well is deep enough.
Also from the result of long term dynamical evolution of embedded stellar system, we expect a final quasi-equilibrium 
state of the system due to the thermal equilibrium with the external potential well acting as a heat bath.

Similar phenomenon has been noticed  from the study of dynamical evolution
of two component stellar systems with relatively large mass ratio. \citet{lee1995,lee2001}
studied the stellar system composed of ordinary solar mass stars and the black holes
of 10 times higher mass. The black holes form compact subsystem through the dynamical
friction in short time scale. As the central density increases the binaries form among black
holes and eventually stops the core collapse. The evolution after the collapse is characterized
by nearly static configuration since the larger stellar system composed of ordinary stars efficiently
absorbs the heat generated by the binaries. Since the embedding system has much larger mass,
the heating does not affect the surrounding system.

\section{Equilibrium models for self-gravitating system embedded in potential well}
Motivated by the expectation of the quasi-equilibrium state in $N$-body simulation results,
we consider equilibrium models for the self-gravitating stellar system embedded in potential well.
Although several equilibrium models of isolated system are known, among which are isothermal, King, Plummer model,
a little attention is given to the equilibrium model of self-gravitating system embedded in potential well. 
Here we propose an equilibrium model of this system based on the argument discussed in Sections 3 and 4. 

A spherically symmetric stellar system with isotropic velocity dispersion satisfies
Jeans equation \citep{bt2008}.
\begin{equation}
\frac{d(\rho \sigma^2)}{dr} = -\rho \frac{d\Phi}{dr}
\end{equation}
And self-gravitating system also satisfies Poisson equation
\begin{equation}
\nabla^2 \Phi = 4\pi G \rho
\end{equation}
If we consider the potential well and assume that
it is deep enough to make the stellar system isothermal with the same temperature of external potential well 
as expected from the simulation results, Jeans equation may be rewritten as
\begin{equation}
\sigma^2 \frac{d\rho}{dr} = -\rho \frac{d\phi_s}{dr}-\rho \frac{d\phi_e}{dr}
\end{equation}
Then, we obtain
\begin{equation}
\rho = \rho_0 e^{-\frac{1}{\sigma^2}(\phi_s + \phi_e)}
\end{equation}
where $\rho_0$ is the central density of embedded system and, $\phi_e$ and $\phi_s$ are the external potential well 
and the potential well of self-gravitating system respectively.
Thus, if we assume spherical symmetry, Poisson equation is
\begin{equation}
\frac{1}{r^2}\frac{d}{dr}(r^2 \frac{d\phi_s}{dr}) =
4\pi G \rho_0 e^{-\frac{1}{\sigma^2}(\phi_s + \phi_e)}
\end{equation}
This becomes the equation of equilibrium isothermal sphere if $\phi_e = 0$.
If we use Plummer potential for $\phi_e$,
\begin{equation}
\phi_e = - \frac{GM_e}{R} \left[ 1 + (\frac{r}{R})^2 \right]^{-1/2}
\end{equation}
and solve the Poisson equation, we obtain the density profile of equilibrium model
embedded in the Plummer potential well.
For solving this equation, we use a normalized length $\xi$
\begin{equation}
\xi = r \left(\frac{4\pi G \rho_0}{\sigma^2}\right)^{1/2}
\end{equation}
where we have two parameters to be set: central density $\rho_0$ and isothermal velocity dispersion $\sigma^2$ 
of embedded self-gravitating system. These two parameters can usually be determined by observation.

\begin{figure}
\center
\epsfig{figure=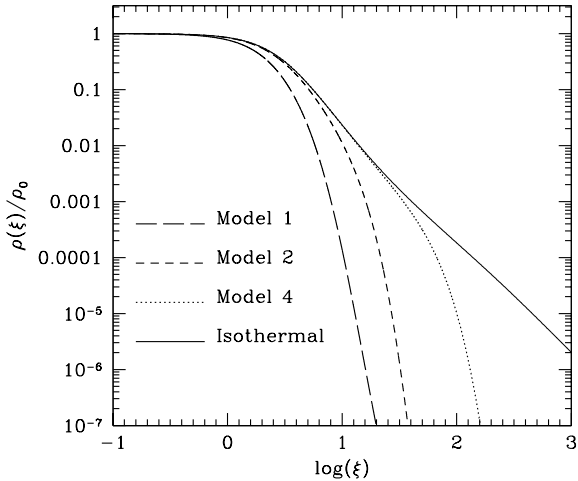, height=0.47\textwidth, width=0.47\textwidth}
\vspace{-3mm}
\caption{The equilibrium density profile of Model1 (lon-dashed), Model2 (dashed) and Model4 (dotted).
The isothermal model (solid) are also shown for comparison.
Model4 and Model1 is the case of the deepest and the shallowest potential well in \tab~\ref{tab1}.
If there is a potential well, it confines the embedded self-gravitating system and makes the density profile of the system 
truncated and deviated from isothermal sphere (thin solid line). 
As potential well becomes deep, the density profile becomes more steep in outer region.
}
\label{fig11}
\end{figure}

In \fig~\ref{fig11}, we show the equilibrium density profiles of isothermal sphere embedded in three different Plummer potential wells 
in Model1, Model2 and Model4 (see \tab~\ref{tab1}). In the figure, thin solid line is isothermal sphere shown for comparison.
Thick green, blue and red solid lines are the equilibrium density profiles for the system embedded in Plummer potentials in Model1, Model2 and Model4.

As shown in \figs~\ref{fig8} and \ref{fig9} in Section 4, 
Model4 collapses due to gravothermal instability and expands later.
Therefore the equilibrium density profile assumed to be in thermal equilibrium with Plummer potential in Model4 (red solid line in \fig~\ref{fig11}) 
is concentrated, however the outer part is truncated due to the potential well confining the system, 
which is in contrast to the case of isothermal sphere (thin black solid line).
Model1 whose equilibrium density profile is shown with green solid line in \fig~\ref{fig11}, 
is gravitationally unbound and the total energy is greater than 0.0 (see \fig~\ref{fig9}) since the deep 
external potential heats the central system and increases its velocity dispersion (i.e. kinetic energy). 
The system is similar to the ideal gas. Thus core collapse would not occur.
Since the potential well confines the outer part of the embedded system and increases the velocity dispersion, 
the central region expands and outer boundary shrinks when it reaches the equilibrium. 
Model2 whose equilibrium density profile is shown with blue solid line in \fig~\ref{fig11}, is
the intermediate case and loosely bound by gravity as the total energy is less than, 
but close to 0.0 (see \fig~\ref{fig9}). As shown in \fig~\ref{fig8}, 
core collapse does not occur until the simulation stops at $T=78T_{rh,i}$, although it might occur after very long time.
Since the potential well of Model2 is not as deep as that of Model1, the outer boundary is not declined
as sharply as Model1. Also the central region does not expand as much as Model1 because 
the increase of velocity dispersion due to the heating by the external potential is not as significant as Model1.  

\begin{figure}
\center
\epsfig{figure=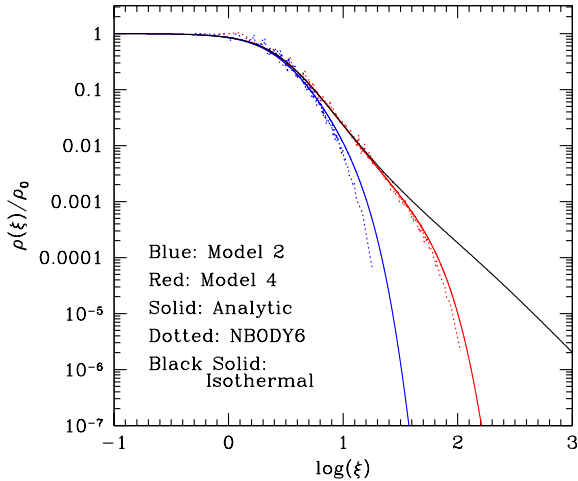, height=0.47\textwidth, width=0.47\textwidth}
\vspace{-3mm}
\caption{
The comparison between the equilibrium density profiles from analytic model (dotted line) and $N$-body model (solid line), 
for Model2 (blue) and Model4 (red).
All density profiles are normalized to the central density, and radius is properly scaled as explained in the text.
Although the equilibrium density profiles from analytic and $N$-body model do not perfectly agree especially at the center, 
the overall shapes are similar. 
}
\label{fig12}
\end{figure}

In \fig~\ref{fig12} we compare the equilibrium density profiles from analytic model (dotted line) and $N$-body model (solid line),
for Model2 (blue) and Model4 (red). Final quasi-equilibrium density profile from $N$-body simulation (black solid lines among
the density profile snapshots of Model2 and Model4 in \fig~\ref{fig10} were compared with equilibrium density profiles of
Model2 and Model4 in \fig~\ref{fig11}. All density profiles are normalized to the central density, and radius is scaled as follows.
We estimate the $\rho_0$ and the velocity dispersion assumed to be isothermal, from density and velocity dispersion profiles
(black lines) in \fig~\ref{fig10}, and calculate the corresponding core radius of isothermal sphere 
$\sqrt\frac{\sigma^2}{4\pi G \rho_0}$ for Model2 and Model4, where $\sigma$ is the radial velocity dispersion estimated 
from $N$-body simulation.
Then we rescale radius of $N$-body density profile by multiplying initial half-mass radius (i.e. recall that
the radius in \fig~\ref{fig10} is scaled by initial half-mass radius), and multiply the corresponding core radius.
Now $N$-body density profile radius is the same $\xi$ as one in \fig~\ref{fig11}.
Although the discrepancy between analytic and $N$-body models is seen at the center possibly due to the small number of 
particles in $N$-body simulation, analytic and $N$-body model show the consistent result.  

\section{Discussion}
Using simple models we studied the effect of surrounding potential well to the surrounded self-gravitating
system, simulated the dynamical evolution of the system and proposed equilibrium models. 
In the following, we summarize and discuss the result.

\subsection{The role of potential well surrounding isothermal self-gravitating system}
We approximate Plummer potential to harmonic potential near the center by Taylor expansion and investigate its effect to 
the isothermal self-gravitating system at the center of the potential well. As the external potential becomes deeper 
compared with that of the embedded self-gravitating system, the external potential behaves like a conducting rigid wall
which permits the heat exchange and conserves the temperature of self-gravitating system.
If the potential depth ratio $A$ is smaller than 1.5, there is a minimum dimensionless energy $\lambda_c$ in $\lambda-\eta$ plane, below
which the system has no equilibrium condition. However if $A$ is greater than 1.5, there is a maximum $\lambda_c$, beyond 
which the system can not be in equilibrium. As the potential depth ratio becomes large, the density contrast 
$D=\frac{\rho_c}{\rho(R)}$ at $\lambda=\lambda_c$ becomes close to 32.1, which is the value for isothermal
sphere enclosed by conducting wall with dimensionless temperature $\eta_c=2.52$.

Thermodynamical description of self-gravitating system is useful for understanding the global evolution of the system. 
Recently similar works in Section 2 are done using a general functional \citep{tsallis1988} which gives
the polytrope with index $n$ \citep{chavanis2002a,chavanis2002b,ts2002,ts2003a,ts2003b}. Isothermal sphere and Plummer model correspond to the polytrope with $n \rightarrow \infty$ and $n=5$ respectively.
However, the maximization of {\it Tsallis functional} at fixed mass and energy is a condition
of dynamical stability rather than thermodynamical stability  \citep{chavanis2006a}. In this context,
polytropic distribution is justified as a particular steady solution of the collisionless Boltzmann equation. 
Furthermore, in this dynamical interpretation, {\it Tsallis functional} is not an entropy.   

Strictly speaking, the self-gravitating system does not have the thermodynamic limit where usually particle number 
$N \rightarrow \infty$ and volume $V \rightarrow \infty$, keeping $N/V$ constant \citep{devega2002a}. 
However, de Vega \& S\'anchez \citep{devega2002a,devega2002b} found the `dilute' 
thermodynamic limit (particle number $N \rightarrow \infty$ and volume $V \rightarrow \infty$, keeping $N/V^{1/3}$ constant) 
where energy, entropy, the free energy are extensive. This study provides a justification of taking thermodynamical approach to 
describe the self-gravitating system, which is useful to understand important physics using much less expensive computational 
resource than numerical simulation.

\subsection{The effect of potential well to the dynamical evolution of the embedded self-gravitating system}
We generate self-gravitating stellar system and surround it using external Plummer potential. 
NBODY6 simulates the evolution of the system and shows the consistent results with the argument in Section 3.
The potential well retards the relaxation process by heating the embedded stellar system and increasing its velocity 
dispersion. Thus if the embedded system has a positive energy $E$, it behaves like an ideal gas and does not experience
the gravothermal catastrophe. On the other hand, the system with negative energy eventually experiences core collapse 
by gravothermal catastrophe, although core collapse time of the system is larger than that of isolated stellar system. 
It is because, as the kinetic energy of stars interacting with the potential increases, the system becomes loosely gravitationally
bound and $T_{rh,i}$ of the embedded system increases (see \eq~\ref{rlxtime}).

The evolution of Lagrangian radii of our model shows that the deep external potential makes the embedded system gravitationally 
unbound and core collapse does not occur as seen in \fig~\ref{fig8}. 
From the energy exchange between the embedded system
and surrounding potential well as seen in \fig~\ref{fig9}, it is suggested that the potential well heats the embedded system and 
increases its kinetic energy. Therefore the total energy of embedded system can be positive if the potential well is deep enough.

From the time evolution of density and velocity dispersion profiles of the embedded system in the external potential well, 
it is very likely that the embedded system is in thermal equilibrium with the potential well.
In deep potential well, we show that the velocity dispersion profile becomes isothermal due to the heating by the potential well.
The inner part of density profile becomes less concentrate and the outer part becomes steeper than the 
isolated system. In the shallow potential well, we see that core collapse occurs, and the velocity dispersion profile is 
not isothermal and has a gradient.

These simulations are based on simple model where the possible interaction between central system and surrounding 
potential well is not considered. 
In order to understand how the embedded system co-evolves with potential well, we need to implement 
the potential well using large number of stellar particle instead of fixed potential function. However, 
for the deep potential well, our simple approach with fixed potential would be a good approximation.

\subsection{Equilibrium configuration for self-gravitating system embedded in potential well}
Based on the conclusion that if the potential well is a heat bath, the embedded self-gravitating system is in isothermal 
equilibrium with the potential well, we derived the equilibrium density profiles of self-gravitating system embedded in potential well by 
solving Jeans equation and Poisson equation. These equilibrium density profiles are similar to $N$-body simulation results.

Especially these equilibrium models are physically motivated by radial distribution profile of galaxies in cluster, 
which is often modelled by King profile \citep{king1966}. King model is a good fit for distribution of galaxies near the cluster core.
However King model looses stars by tidal energy cut-off. This picture is unrealistic in the case of galaxies in
cluster, which are embedded in the deep potential well of cluster dark matter halo. Galaxies in cluster are bound and hard to escape from 
cluster potential well. Our model can be more realistic description for galaxy clusters. 
While the King profile drops at the outer part due to tidal energy cut-off, our model profile drops because 
the potential well confines the embedded system and keeps stars from escaping out.

\section*{Acknowledgments}

This research was supported by KRF grant No. 2006-341-C00018. 
The computation was done on the GPU computer provided by the
grant from the National Institute for
Mathematical Sciences through the Engineering Analysis Software Development program.
We wish to thank P. Berczik, S. Aarseth and R. Spurzem for helping us in porting the GPU version of  
NBODY6 code.

\end{document}